\documentclass[fleqn,usenatbib]{mnras}
\usepackage[T1]{fontenc}
\DeclareRobustCommand{\VAN}[3]{#2}
\let\VANthebibliography\thebibliography
\def\thebibliography{\DeclareRobustCommand{\VAN}[3]{##3}\VANthebibliography}

\usepackage{amsmath}
\usepackage{graphicx}
\usepackage{txfonts}
\usepackage{tablefootnote}
\usepackage{threeparttable}
\usepackage{subcaption}
\usepackage{placeins}

\title[]{Classification of 3 accreting binaries with VLT/X-Shooter spectra}

\author[]{
T. Bouchet,$^{1}$\thanks{E-mail:tristan.bouchet@obspm.fr}
S. Chaty,$^{2,1}$\thanks{E-mail:sylvain.chaty@u-paris.fr}
F. Fortin$^{2}$\thanks{E-mail:fortin@apc.in2p3.fr}
and J. A. Tomsick$^{3}$
\\
$^{1}$Université de Paris and Université Paris Saclay, CEA, CNRS, AIM, F-91190 Gif-sur-Yvette, France\\
$^{2}$Université Paris Cité, CNRS, Astroparticule et Cosmologie,  F-75013 Paris, France\\
$^{3}$Sciences Laboratory, 7 Gauss Way, University of California, Berkeley, CA 94720-7450, USA
}

\date{}
\pubyear{2022}

\begin{document}
\label{firstpage}
\pagerange{\pageref{firstpage}--\pageref{lastpage}}
\maketitle

\begin{abstract}
Since its launch, the \emph{INTErnational Gamma-Ray Astrophysics Laboratory} (\emph{INTEGRAL}) satellite has discovered hundreds of X-ray sources, many of which lack proper classification. This mission also led to the discovery of new categories of high mass X-ray binaries (HMXB). We use the spectra of the X-Shooter instrument at the Very Large Telescope (VLT) of the European Southern Observatory (ESO) to better understand the nature of 3 accreting binaries (IGR J10101-5654, IGR J11435-6109 and IGR J12489-6243) discovered by \emph{INTEGRAL}. We mainly focused on the lines and continuum from the X-Shooter spectra. We used atlases to constrain the nature of the sources and also complemented the spectra with measurements taken by \emph{Spitzer} and the \emph{Wide-field Infrared Survey Explorer} (\emph{WISE}) in infrared, and parallaxes from \emph{Gaia} for the distances. We determined the nature of each binary system: a BeHMXB system with a companion star of spectral type B0.5Ve with peculiar carbon emission for IGR J10101-5654 and IGR J11435-6109, and a CV system with an evolved K star (K0IV-K2IV) for IGR J12489-6243. We also estimated some geometrical parameters of the decretion disk and neutron star's orbit in the case of IGR J11435-6109.
  {}
\end{abstract}


\section{Introduction}
In the 1960's the first X-ray sources of extra-solar origin have been detected and quickly identified as accreting binary systems \citep{giacconi1962}. 
Afterwards other surveys were done in the soft X-ray domain (< 20 keV) to carry on the study of those sources (\emph{Uhuru} in 1970, \emph{ROSAT} in 1990 and \emph{ASCA} in 1993). In the last two decades more missions have been launched in various energy bands, for instance \textit{Chandra} and \textit{XMM-Newton} in soft X-ray, and \textit{INTEGRAL}, which has provided hard X-ray images of the Galactic plane. This in turn led to the discoveries of many more X-ray sources and our understanding of those peculiar binaries has greatly improved since then \citep{integral2019}.\\

X-ray binaries are systems with a compact object accreting matter attracted from a companion star, which leads to the emission of high energy radiation mainly detected in the X-ray domain or above. Here we report on the study of two types of accreting binaries: High Mass X-ray binaries (HMXB), which consist of a high mass (\(M \gtrsim 10 M_\odot\)) companion, generally a O or B-type star, while the compact object is a black hole or neutron star; the other type consists of cataclysmic variables (CV), involving a low mass star (\(M \lesssim 1 M_\odot\)) with a white dwarf as the compact object. \\ 

Furthermore, we can distinguish 2 sub-categories of HMXB depending on the accretion process of the compact object. When the companion star is a main sequence Be, its fast rotation will shed matter around itself, forming a decretion disk; the compact object will then accrete matter periodically due to its eccentric orbit (\citet{Okazaki_2001}, see also \citet{Martin_2021} for aperiodic Type I outbursts). This process can sometimes be detected in the evolution of the X-ray spectrum, where the luminosity will be modulated by the rotation of the neutron star and its orbit around the Be star.  The decretion disk will also be detectable in the spectrum at lower energy, in the near-infrared and optical domain, and will appear as emission lines with peculiar shapes that depends on the shape and inclination of the disk. This disk is also highly variable in time and contributes to the continuum, which significantly affects the brightness of the source.
In the case of a supergiant companion, the compact object will accrete a fraction of the wind coming from the companion star \citep{darkside2018, CHATY20132132}. These sources show very little variability in X-ray, NIR or at optical wavelengths. When the star or the binary is surrounded by cold material it can lead to the emission of forbidden lines, which can be detected near optical wavelength. It can sometimes be attributed the stellar wind itself \citep{forbidden_line_wind}.
The study of those binary systems have been of particular interest in the recent years with the novel observations of gravitational waves. Indeed HMXB are a common stage in the life of high-mass binaries, and understanding this stage better can help to constrain parameters of latter stages that eventually lead to the merging of compact objects, or lack thereof \citep{ligo}.\\

In CV the accretion is done through Roche lobe overflow, meaning that the companion star will fill its Roche lobe and lose matter through the L1 Lagrange point. The white dwarf will then accrete matter through two different processes: with an accretion disk if the magnetosphere is weak, or through its poles if its magnetosphere is too strong for a disk to form around it, in this case the system is called a polar. There also exists an in-between case where a disk can form outside the magnetosphere called an intermediate polar \citep{accretion_process}.\\

In our case, the discoveries of the 3 X-ray sources were made with the \emph{INTEGRAL} satellite observing between 20 keV and 10 MeV, but the high energy observations alone don't allow us to understand the physical processes at play in detail, especially the nature of the companion star. Those sources were also previously studied at near-infrared wavelength but for each of these sources some questions remain that we aim to answer: IGR J10101-5654 has an unclear presence of forbidden lines and double lines despite being identified as a sgB[e], IGR J11435-6109 has a similar spectrum to IGR J10101-5654 although it was found to be a BeHXMB, and IGR J12489-6243 has a surprising lack of molecular bands in the K band despite being identified as a CV system. This will be detailed in section \ref{analysis}.\\

Hence we need to study the spectra of those sources on a broader range of wavelength, which is possible with the X-Shooter instrument at the VLT in Chile. We aim to understand the nature of those sources using spectroscopy. We will first describe the observations made and how the data were reduced, and then make a detailed analysis of the spectra of each object.\\

\section{Observations and reduction}
\subsection{Observations}
The observations of the sources were carried out by the X-Shooter spectrograph at the VLT in February of 2019. This instrument has 3 different arms that can observe in the ultraviolet (UVB), optical and near-infrared (NIR) domains, covering between 300 and 2480 nm. The observation of the 3 sources is summarized in Table \ref{table:observations_target}, where we give observation times, exposure times and airmass. For all 3 sources, the slit dimensions and spectral resolutions were 1.0"x11" in the UVB (R=5400), 0.9"x11" in VIS (R=8900) and 0.9"x11" in NIR (R=5600). Slit losses were accounted for with atmospheric dispersion compensators in UVB and VIS and by a tip-tilt mirror in NIR.

\subsection{Spectra processing}
\label{spectra_processing}
The reduction of the data was done with the X-Shooter pipeline of the ESOReflex software\footnote{https://www.eso.org/sci/software/pipelines/}. It combines the different spectral orders of the spectrum and takes into account the bias, dark, sky subtraction and airmass correction. The wavelength calibration and normalization is done using calibration lamps (ThAr for VIS and UVB, Hg/Ar/Ne/Xe for NIR). The flux calibration uses standard star spectra taken at the same time (see Table \ref{table:observations_telluric_std}).\\

Sky subtraction was done with the nodding procedure of the ESOReflex pipeline except for IGR J10101-5654, which had another source polluting the spectrum. Therefore we had to subtract the sky emission manually which led to the UVB and some of the visible part of the spectrum to become negative since we could not totally get rid of the polluting star.

\begin{table*}
\caption{X-Shooter observations - the 3 targets of interest}             
\label{table:observations_target} 
\begin{threeparttable}
\centering
\begin{tabular}{cccccccc}
\hline\hline
 Sources              & RA J2000             & DEC J2000            & Domain               & Observation date (UTC)   & Exp. time            & \multicolumn{2}{c}{Airmass}       \\
\multicolumn{1}{l}{} & \multicolumn{1}{l}{} & \multicolumn{1}{l}{} & \multicolumn{1}{l}{} & \multicolumn{1}{l}{}     & \multicolumn{1}{c}{(s)} & \multicolumn{1}{l}{Start} & End   \\ \hline
&&        & UVB                  & 2019-02-07T01:21:58.9639 & 1776.0                & 1.805                     & 1.53  \\
     IGR J10101-5654      & 10:10:11.87       & -56:55:32.1               & VIS                  & 2019-02-07T01:21:55.815  & 4752.0                 & 1.805                     & 1.53  \\
                     &                      &                      & NIR                  & 2019-02-07T01:21:50.605  & 2112.0                 & 1.806                     & 1.536 \\ \hline
&  && UVB                  & 2019-02-15T03:08:58.0911 & 1824.0                 & 1.613                     & 1.436 \\
IGR J11435-6109      & 11:44:00.30       & -61:07:36.5                         & VIS                  & 2019-02-15T03:08:55.347  & 4800.0               & 1.613                     & 1.436 \\
                 &                      &                      & NIR                  & 2019-02-15T03:08:50.136  & 2112.0                 & 1.613                     & 1.44  \\ \hline
&  & & UVB                  & 2019-02-19T02:20:01.3968 & 1776.0                 & 2.218                     & 1.839 \\
IGR J12489-6243      & 12:48:46.422       & -62:37:42.53                  & VIS                  & 2019-02-19T02:19:58.102  & 4752.0               & 2.219                     & 1.838 \\
             &                      &                      & NIR                  & 2019-02-19T02:19:52.882  & 2112.0                & 2.22                      & 1.847 \\ \hline

\multicolumn{1}{l}{} & \multicolumn{1}{l}{} & \multicolumn{1}{l}{} & \multicolumn{1}{l}{} & \multicolumn{1}{l}{}     & \multicolumn{1}{l}{} & \multicolumn{1}{l}{}      &       \\
\multicolumn{1}{l}{} & \multicolumn{1}{l}{} & \multicolumn{1}{l}{} & \multicolumn{1}{l}{} & \multicolumn{1}{l}{}     & \multicolumn{1}{l}{} & \multicolumn{1}{l}{}      &      
\end{tabular}
\end{threeparttable}
\end{table*}

\begin{table*}
\caption{X-Shooter observations - Telluric standard stars used for the flux calibration}             
\label{table:observations_telluric_std} 
\begin{threeparttable}
\centering
\begin{tabular}{cccccccc}
\hline\hline
 Sources              & RA J2000             & DEC J2000            & Domain               & Observation date (UTC)   & Exp. time            & \multicolumn{2}{c}{Airmass}       \\
\multicolumn{1}{l}{} & \multicolumn{1}{l}{} & \multicolumn{1}{l}{} & \multicolumn{1}{l}{} & \multicolumn{1}{l}{}     & \multicolumn{1}{c}{(s)} & \multicolumn{1}{l}{Start} & End   \\ \hline
Hip033998      & 07:03:14.08224      & -11:08:15.468                  & VIS                  & 2019-02-07T00:23:38.703  & 312.0            &1.215                    & 1.206 \\
               &                      &                      & NIR                  & 2019-02-07T00:23:41.8966  & 168.0               & 1.215                     & 1.208 \\ \hline
Hip033998      & 07:03:14.08224      & -11:08:15.468                  & VIS                  & 2019-02-15T00:25:58.120  & 192.0            &1.127                  & 1.122 \\
              &                      &                      & NIR                  & 2019-02-15T00:14:09.9469  &  120.0              & 1.153                     & 1.149 \\ \hline
Hip031962     & 06:40:47.62104      & -47:37:15.168                 & NIR                  & 2019-02-19T05:22:22.1691  & 180.0            & 1.595                 & 1.604 \\ \hline

\multicolumn{1}{l}{} & \multicolumn{1}{l}{} & \multicolumn{1}{l}{} & \multicolumn{1}{l}{} & \multicolumn{1}{l}{}     & \multicolumn{1}{l}{} & \multicolumn{1}{l}{}      &       \\
\multicolumn{1}{l}{} & \multicolumn{1}{l}{} & \multicolumn{1}{l}{} & \multicolumn{1}{l}{} & \multicolumn{1}{l}{}     & \multicolumn{1}{l}{} & \multicolumn{1}{l}{}      &      
\end{tabular}
\end{threeparttable}
\end{table*}

\begin{figure*}
\centering
\includegraphics[scale=0.6]{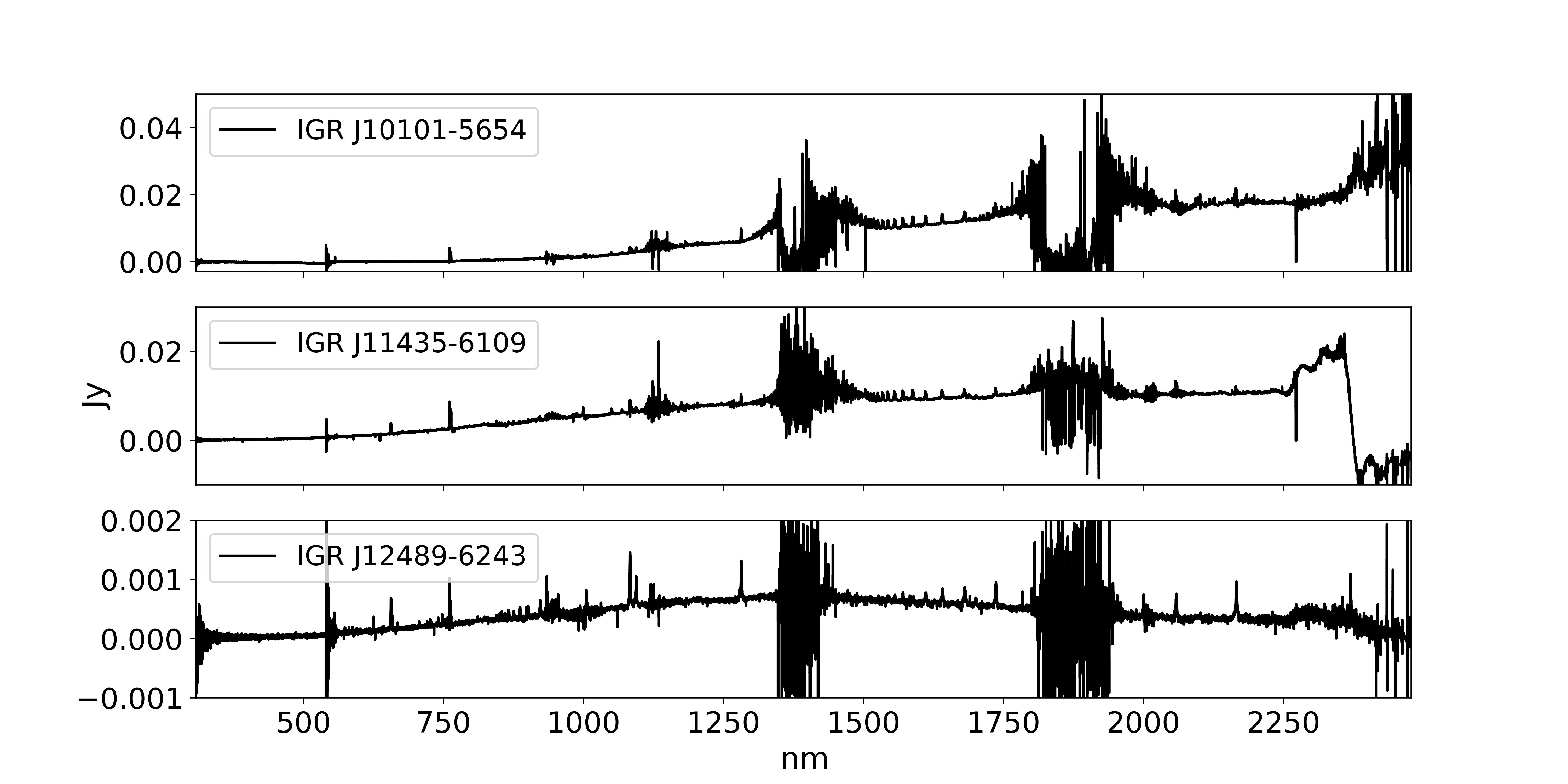}
\caption{Full X-Shooter spectra after reduction, some residual absorption features from the atmosphere are still visible}
\label{fig:full_spectra}
\end{figure*}

To correct for telluric absorption, we used the Molecfit software \citep{molecfit1,molecfit2} which creates a transmission curve of the atmosphere using the theoretical molecular absorption and parameters depending on the local weather. Unfortunately the regions near two important lines at 2 $\mu m$ could not be corrected; these regions correspond to $CO_2$ absorption bands. The final spectra can be seen in Fig.\ref{fig:full_spectra}.\\

We compared the photometric flux measured by the Gaia mission (at 504, 582, 762 nm) and 2MASS (at 1240, 1650 nm) with the flux of our spectrum at the same wavelengths. IGR J11435-6109 and IGR J12489-6243 have compatible flux, while IGR J10101-5654 shows a discrepancy of around 50\%, which could be explained by the temporal variation of the emission, previously noticed in X-ray \citep{Tomsick_2008}.

\subsection{Complementary data}
We also used observations from other missions to complete our analysis, in particular when modeling the continuum.\\

For near-infrared data above 2.4 $\mu m$ we used photometric measurements from the \emph{Spitzer} and \emph{WISE} missions, which give the magnitude of the sources in the L, M, N and Q bands. For the distances, we relied on the last release available of the \emph{Gaia} astrometric mission Gaia EDR3 \citep{gaia_main,gaia_dr3}. Distances for the sources were computed by \citet{Bailer_Jones_2021}, who used the parallaxes along with parameters that affect the parallax measurement such as the magnitude and the color of the star, and a distance prior based on the density of stars in the galaxy.

This information, along with X-ray fluxes and luminosities, is summarized in Table \ref{table:2}.

\begin{table*}
\caption{Complementary data for the 3 sources}
\begin{threeparttable}
\centering
\begin{tabular}{cccccccc}
\hline \hline 
Sources         & L band & M band & N band & Q band &  Distance & X-ray flux & X-ray luminosity \\
                & mJy    & mJy    & mJy    & mJy    & kpc            & $10^{-12}$ erg/cm2/s  &  $10^{33}$ erg/s            \\ \hline
                \\
IGR J10101-5654 &  31.1 $\pm$ 0.6  \tnote{a}    &   25.3 $\pm$ 0.4  \tnote{b}   &  8.2 $\pm$ 0.3  \tnote{c}    &  \O     &   $4.5^{+1.1}_{-1.2}$ \tnote{*}  &  $2.03^{+0.35}_{-0.25} $  &  $4.9^{+3.9}_{-1.9}$                \\
\\
IGR J11435-6109 &  13.3 $\pm$ 0.3 \tnote{a}    &  9.5 $\pm$ 0.2   \tnote{b}   &  3.3 $\pm$ 0.2   \tnote{c}   &   3.1 $\pm$ 0.6   \tnote{d}   &  $7.9^{+1.4}_{-1.2}$      &   $9.1^{+4.8}_{-1.7}$  &  $67^{+30}_{-18}$            \\
\\
IGR J12489-6243 &   0.63 $\pm$ 0.03 \tnote{e}   &  0.43 $\pm$ 0.06  \tnote{f}    &  \O      &     \O   &       $2.6^{+0.8}_{-0.5}$  &   $0.54^{+0.25}_{-0.11}$      & $0.42^{+0.67}_{-0.08} $ \\ \hline 

\end{tabular}

\begin{tablenotes}[para] 
\item [a] at 3.3526 $\mu m$,
\item [b]at 4.6028 $\mu m$, 
\item [c]at 11.5608 $\mu m$, 
\item [d]at 22.0883 $\mu m$, 
\item [e]at 3.6 $\mu m$, 
\item [f]at 4.5 $\mu m$
\\
\item[*] this distance is uncertain due to the high astrometric noise parameter of the associated \emph{Gaia} measurements
\item X-ray corresponds to the integrated flux between 0.3 and 10 keV and are taken from \citet{Tomsick_2008} and \citet{Tomsick_2012}.
\item The corresponding luminosities were computed using the distances and X-ray flux\\
\item Distances from \citet{Bailer_Jones_2021}
\end{tablenotes}

\end{threeparttable}
\label{table:2}
\end{table*}

\section{Analysis}
\label{analysis}
After the spectra have been properly reduced and calibrated, we can start to analyze their two main attributes:
\begin{enumerate}
      \item the lines, which can tell us about the chemical composition of the stellar photosphere and the environment of the binary, as well as the type of environment (motion, presence of a disk of dust...)
      \item the continuum, which corresponds to the Spectral Energy Distribution (SED) that relates to the source(s) of luminosity in the binary system and the extinction induced by the material between us and the source. In our case we extracted the continuum by making a logarithmic binning of the X-Shooter spectrum.
\end{enumerate}

\begin{figure*}
\centering
\includegraphics[scale=0.5]{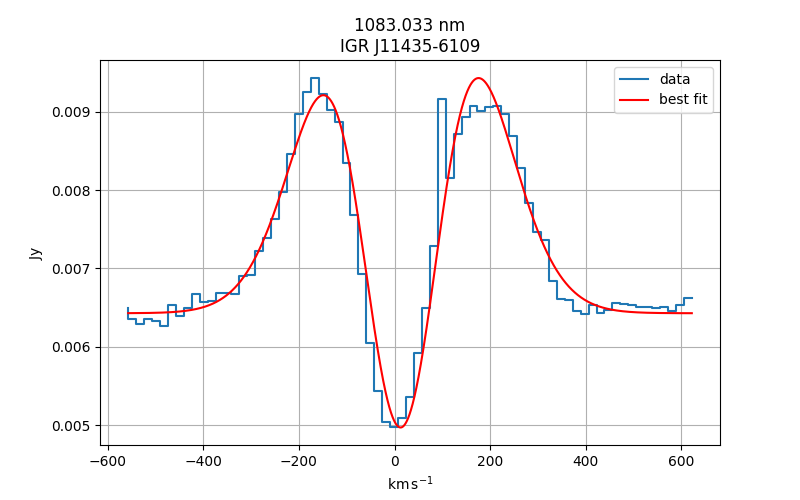}
\caption{Example of a shell profile of a Helium line, the absorption component falls below the continuum and is usually thinner than the emission}
\label{fig:auto_abs}
\end{figure*}

\begin{figure*}
\centering
\includegraphics[scale=0.5]{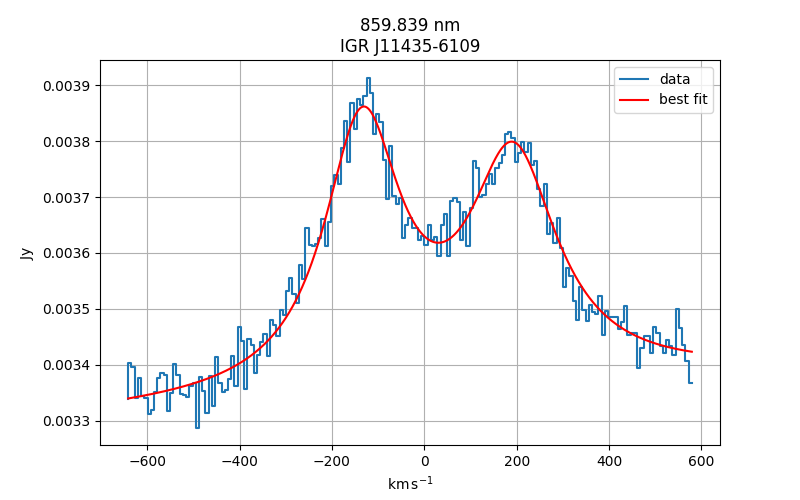}
\caption{Example of a double-peaked profile of a Hydrogen line, the depression in the center is due to the shape of disk and its inclination}
\label{fig:double_pic}
\end{figure*}

After identifying the lines in each spectrum, we fitted them with a Gaussian or Lorentzian profile when single peaked. Profiles where an absorption component is close to or lower than the local continuum were considered as a shell profile and fitted by a sum of an absorption and emission component (see Fig.\ref{fig:auto_abs}). Otherwise double peaked profile were fitted by a sum of 2 emissions (see Fig.\ref{fig:double_pic}), in this case the relevant parameters are the separation between the two peaks noted $\delta v_p$ and the ratio of intensity between the two peaks, noted $V$\slash$R$ ($V$, for \emph{Violet}, being the intensity of the blueshifted peak and $R$ for the redshifted peak). The $V$\slash$R$ ratio can give an indication on the asymmetry of the disk and can be particularly useful to follow the evolution of the shape of the disk through time when several spectra are available \citep{VR_ratio}. In our case the uncertainty on this ratio is always higher than 50\% so it is of limited interest.\\ 

As for the continuum, the fitting process was done using the least-square method with the lmfit Python library and the high sky absorption regions were removed. For all fits we took the \emph{Gaia} distances found previously.

\subsection{Diffuse interstellar bands}
An attempt was made to find a value for the reddening of the spectra $E(B-V)$ using the diffuse interstellar bands (DIBs). Those bands are absorption features that can be found in many spectra independently of the nature of the source observed: indeed their origin is the interstellar medium. Several authors found a linear correlation between their equivalent width and the reddening of the type \(EQW = a \ E(B-V) + b\) with $a$ and $b$ parameters depending on the DIB used \citep{Groh_2007,dibs_absorption, DIBs862nm}.\\

\begin{figure}
\centering
\includegraphics[scale=0.3]{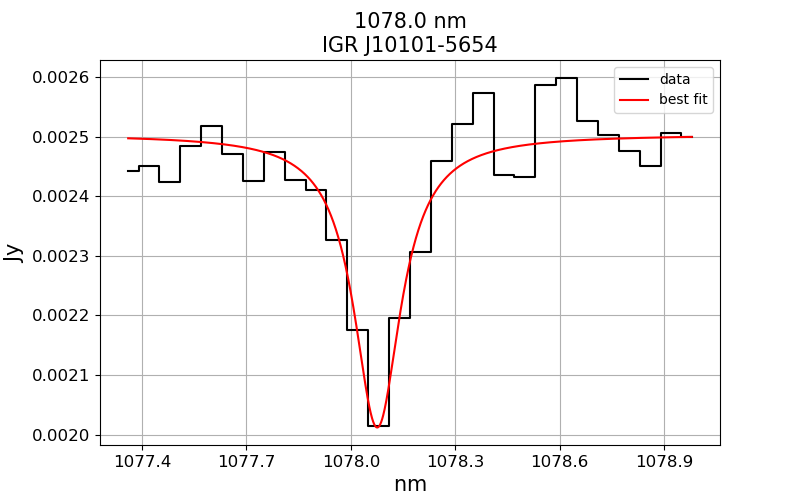}
\caption{Diffuse Interstellar Band at 1078 nm, fitted with a Gaussian}
\label{fig:dib_1078}
\end{figure}

We only observed those DIBs for the 2 Be sources, and only 4 different DIBs were used . We fitted them with Gaussians (see Fig. \ref{fig:dib_1078}) and evaluated their equivalent width to deduce the reddening (see Table \ref{table:dibs}). For the 3 DIBs at 1179 and 1317 nm, the uncertainties are too high ($\gtrsim$ 40\%) to be of interest, but the DIB at 1078 nm has lower uncertainties. For IGR J11435-6109 in particular, the reddening found with the DIB at 1078 nm  ($E(B-V) = 1.7 \pm 0.4$) is compatible with the value found with the dust extinction map of \citet{Schlafly_2011} ($E(B-V) = 1.8$). The reddening found with the 862 nm DIB ($2.3 \pm 0.7$) is also compatible within the error margins.\\

This information could be useful in the future when the reddening cannot be properly estimated by other means and DIBs are observed in the spectrum. In the rest of this paper we used other methods to find the reddening, namely a dust extinction map and the Balmer decrement in the case of IGR J12489-6243.\\

\begin{table*}

\caption{Reddening from the EQW of several DIBs. The last line, given for comparison, corresponds to the value found with a dust extinction map}
\label{table:dibs}
\centering
\begin{tabular}{ccccc}
\hline \hline 
Wavelength & \multicolumn{2}{c}{IGR J11435-6109} & \multicolumn{2}{c}{IGR J10101-5654}                 \\
                & EQW             & E(B-V)    & EQW                               & E(B-V) \\ \hline
                
862 nm   \citep{DIBs862nm}       & 1079 $\pm$ 116 m\AA      & 2.3 $\pm$ 0.7    & \multicolumn{1}{c}{1045 $\pm$ 253 m\AA}  & 2.4 $\pm$ 0.4   \\

1078 nm   \citep{Groh_2007}       & 253 $\pm$ 27 m\AA      & 1.7 $\pm$ 0.4    & \multicolumn{1}{c}{376 $\pm$ 42 m\AA}  & 2.5 $\pm$ 0.6   \\
1179 nm  \citep{dibs_absorption}       & 266 $\pm$ 35 m\AA      & 3.4 $\pm$ 1.6      & \multicolumn{1}{c}{507 $\pm$ 174 m\AA}  & 6.8 $\pm$ 4.4   \\
1317 nm  \citep{dibs_absorption}       & 903 $\pm$ 145 m\AA     & 2.8   $\pm$ 1.0    & \multicolumn{1}{c}{1214 $\pm$ 341 m\AA} & 3.8  $\pm$ 1.8  \\
&  && \multicolumn{1}{c}{ }  & \\
\citet{Schlafly_2011} &  &  1.8 & \multicolumn{1}{c}{}  & 4.4 \\\hline 

\end{tabular}

\end{table*}

\subsection{The 3 accreting binaries}
\subsubsection{IGR J10101-5654}
\paragraph{\emph{Previous results}}

This source was discovered in 2006 by \citet{10101_decouverte}. Its counterpart was found at visible wavelength by \citet{10101_masetti} the same year and identified as a HMXB. According to the authors, the equivalent width of the H$\alpha$ line hints towards a giant or subgiant companion star.
\citet{Tomsick_2008} also found a variability in the X-ray flux, which could mean an eccentric orbit for the compact object. A study of the infrared spectrum then identified the source as a sgB[e] HMXB because of the presence of forbidden emission line \citep{Coleiro_2013}. They could also fit the continuum with a 20,000 K B star and a second spherical black body at 1016 K to take into account the circumstellar dust.\\

\paragraph{\emph{Our results}}

In our X-Shooter spectra, we found hydrogen lines in emission: they are single peaked profile for the Balmer series but double peaks for the Paschen and Brackett series in infrared. The double peaked lines are well adjusted by the sum of two Lorentzians, which would mean they originate from a disk. Some single-ionized and neutral metals are also seen as double peaks or as simple emission lines, and a single FeII line was detected in absorption. The value of the fit parameters are summarized in Table \ref{table:10101_double} and \ref{table:10101_simple}.\\

We do not find any P-Cygni profile that would originate from the wind of the supergiant. We also found a single HeI line at 1083 nm which presents a shell profile (a broad emission line combined with an absorption line) with a FWHM of $140 \pm 2 km/s$. The shape is compatible with a high-inclination ($i>70^\circ $).

In our spectrum we do not detect the two forbidden emission lines of FeII found by \citet{Coleiro_2013} although we do observe what could be another [FeII] line at 1676.8 nm. The inconsistencies between the supposed forbidden lines could be explained by a high time-variability of the source, but this seems highly unlikely, especially when the signal-to-noise ratio is measured at only 2.5 for our line, and estimated at around 3 for the lines of \citet{Coleiro_2013}. We therefore decided to exclude this feature from our analysis.\\

We then compared the emission lines present in the H-band with a stellar atlas for Be star \citep{Chojnowski_2015}, in particular Fig.7 and A22 of their article, and find the most likely stellar type to be a B0.5Vnnep. The 'p' (for peculiar) refers to the presence of a strong double-peaked neutral carbon line at 1689 nm, whose origin is yet unknown. Some of this peculiar feature is hidden by the telluric absorption, but it is clearly observed for the next source (IGR J11435-6109). Because this feature is uncertain, another analysis was done using lines between 1040 and 1090 nm and another spectral atlas \citep{Groh_2007}, in particular Fig.5 of their article, gives a spectral type closer to a B1IVe. In both cases, the companion star seems to be a B0-B1 near the main sequence, and nothing indicates that it could be a supergiant.\\

Because of the temporal variation of the source (as discussed in section \ref{spectra_processing}), we only fitted the X-Shooter data. Therefore we only had enough information to use a simple reddened black body to fit the companion star at a fixed temperature of 28,000 K (corresponding to a B0.5). We find an extinction $A_V$ of $14.7\pm 0.2 \ mag$ and taking into account the uncertainty on the distance, a stellar radius of  $13.6 \pm 3.6\ R_{\odot}$. This value should be taken with care since the distance deduced by \citet{Bailer_Jones_2021} is not necessarily reliable, due to the low S/N ratio associated with the \textit{Gaia} measurements.

\begin{figure*}
\centering
\includegraphics[scale=0.5]{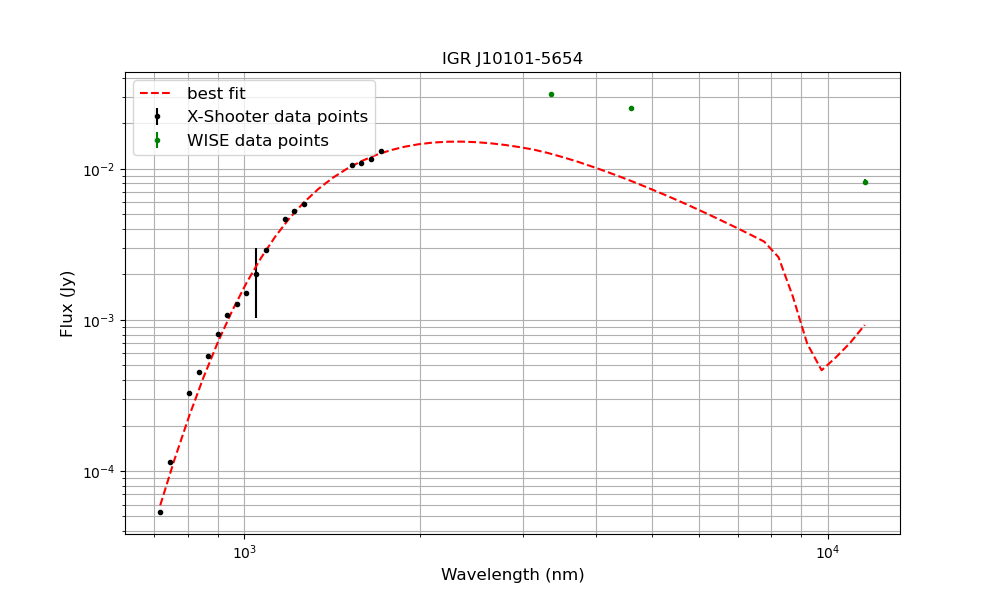}
\caption{Fitting of the SED of IGR J10101-5654, only the X-Shooter data points were used for fitting, the \emph{WISE} data points are here for comparison and show an infrared excess compared to the single black body model.}
\label{fig:sed_10101_cn_xshooter}
\end{figure*}

When we add the \emph{WISE} data points above 2.4 $\mu m$ for comparison, we clearly see a strong infrared excess (see Fig.\ref{fig:sed_10101_cn_xshooter}), which is likely due to circumstellar matter in the form of a disk. We were not able to model the continuum with a full disk model, but the radius found with our simple fit at least gives a maximal radius for the star of 13.6 $R_{\odot}$. This radius, as an upper limit, is more compatible with a main-sequence star than a supergiant, whose radii are typically above 30 $R_{\odot}$ for B stars \citep{radius_stars_book}.

The extinction we found with our model contains both the contribution of the interstellar medium and the circumstellar matter. \citet{Ebv_halpha} show that there is a correlation between the equivalent width of the $H_{\alpha}$ line and the extinction from circumstellar matter. After normalizing the spectrum we found an EQW of $W^{spectrum}=-9.7\ \AA$ for $H_{\alpha}$, and assuming a type B0.5 V companion star, we removed the stellar contribution from the EQW to find $W^{env}=-12.7\ \AA$. Using the correlation given by \citet{Ebv_halpha} we find the reddening induced by the envelope to be $E^{cs}(B-V) = 0.14 \pm 0.04\ mag$, meaning an extinction of $0.4 \pm 0.1\ mag$. After correcting for the presence of the envelope, the extinction from the interstellar medium is then $A_V = 14.3\pm 0.3 \ mag$.\\

We conclude that this HMXB is most likely a BeHMXB rather than a supergiant B[e], since:
\begin{enumerate}
    \item we do not detect any forbidden line feature
    \item the lines are compatible with a main sequence star in two different catalogues
    \item the maximum stellar radius found with a continuum fitting (although this measure is uncertain) is too small for a typical supergiant 
    \item no P-Cygni profiles are seen
\end{enumerate}
The discrepancy between flux measurements, both in X-ray (between \emph{Chandra} and \emph{INTEGRAL}) and in optical (between our observations and \emph{Gaia}) also hints towards a temporal variability of the source which is not compatible with a wind-fed system.\\

The companion star is found to be a B0.5Ve with a radius of at most $13.6 \pm 3.6\ R_{\odot}$.

\begin{table}
\caption{Double peaked emission lines of IGR J10101-5654}
\begin{threeparttable}
\centering
\begin{tabular}{cccccc}
\hline \hline 
Line    & $\lambda_0$ & $\delta v_p$     & $V$\slash$R$         & $v_{rad}$  \\
        & nm       & $km/s$ &     &     $km/s$ \\
\hline
Paschen & 854.53   & 285  &  1.4  &  -21  \\
        & 859.83  &  279  &  1  &   -4       \\
        & 866.502 &  253  &  1.7  &  -27 \\
        & 875.04  &  287.3  &  1.0  &   -9   \\
        & 886.28  &  298.0  &  1.1  &  -22   \\
        &1004.98 &  243.5  &  1.0  &  -17 \\
        & 1093.81 &  230  &  0.8  &   -2 \\
        & 1281.80  &  223.4  &  1.0  &   -5  \\
Brackett &1534.17  &  285.1  &  1.0  &  -11 \\
& 1543.89 &  282.5  &  1.0  &   0 \\
& 1555.64 &  285.1  &  1.0  &   -5\\
& 1570.06 &  289.4  &  1.1  &   -5 \\
& 1588.05 &  275.6  &  0.9  &   -3\\
&   1610.93  &  277.7  &  1.0  &   -3 \\
    &    1640.68  &  265.3  &  0.9  &    2      \\
    &1680.65 &  259.7  &  1.0  &   -3      \\
     & 2165.529 &  223.5  &  1.2  &   -7 \\  
\hline
FeI  & 999.7 &  284.6  &  0.9  &   -2 \\
FeII    & 1050.15 &  313  &  1.3  &    2 \\
FeII    & 1086.35  &  288  &  0.8  &  -15 \\
MgII? & 1091.42 &  378  &  0.3  &   17 \\
CI & 1689.50 &  389.5  & 1.2  & -75\\
MgII & 2143.7 &   \O  & \O  &  \O \\\hline 
\end{tabular}

\begin{tablenotes}
\item $V/R$ is the ratio of intensities of the blushifted peak over the redshifted peak
\end{tablenotes}
\end{threeparttable}
\label{table:10101_double}
\end{table}

\begin{table}
\caption{Simple lines of IGR J10101-5654}
\begin{threeparttable}
\centering
\begin{tabular}{cccccc}
\hline \hline 
Line    & $\lambda_0$ & FWHM     & EQW     & Flux         & $v_{rad}$  \\
        & nm       & $\AA$ &    $\AA$ &    $erg/s/cm^2$   &   $km/s$ \\
\hline
  FeI?      & 906.0 &  1.4  & -1.4  &  1.5e-16  & 25      \\
  FeII      & 917.59   &  1.29  &  0.1  & -4.1e-16  &  4   \\
   CI     &  1068.30 & 6.9 & -0.17 & 1.0e-15  & 11   \\
   CI     &  1069.12 &  5.3    & -0.15      &  8e-16     &  -20  \\
 FeI &  1232.02 &  9.2  & -0.4  &  4.4e-16  & 57 \\
 NI & 1246.12 & 18.6 & -1.3 & 1.7e-15 & -29 \\
 NI & 1246.96 & 13.6 & -0.6 & 5.2e-16 & 45 \\
 MgII & 2137.4 & 14  & -0.31  &  7.0e-16  & 73 \\\hline 

\end{tabular}
\begin{tablenotes}
\item A negative value of EQW indicates an emission feature, and a positive value an absorption feature
\end{tablenotes}
\end{threeparttable}
\label{table:10101_simple}
\end{table}

\subsubsection{IGR J11435-6109}
\paragraph{\emph{Previous results}}

This source was discovered by \citet{11435_decouverte} in 2004. It was then found at lower energy by the \emph{Chandra} satellite and classified as a HMXB \citep{11435_binaire}. The authors also detected the counterpart at infrared and visible wavelength. A more thorough study was done by \citet{Coleiro_2013} on the near-infrared spectrum, and they classified the source as a BeHMXB with a companion star of type B0.5Ve, using the intensity ratio of HeI at 2.05 $\mu m$ and HI at 2.16 $\mu m$. The continuum was fitted in the same way as IGR J10101-5654, with a second black body at 925 K.\\

\paragraph{\emph{Our results}}
We detect all the hydrogen lines in double peak emission (see Table \ref{table:11435_H}), except $H_{\beta}$ which is in shell profile. Helium is only found in its atomic form, either as simple absorption lines or as shell profiles. The fits of the shell profile lines are summarized in Table \ref{table:11435_auto} and indicates a high inclination for the system. The remaining fits for simple lines of the various metals found are summarized in Table \ref{table:11435_simple}.\\

As for the previous source, we focused on the H-band and used the atlas of \citep{Chojnowski_2015} which gave us a spectral type of B0.5Vnnep. This is confirmed by the fact that HeI is seen in absorption at 587 and 667 nm and no HeII lines are detected, typical of early B stars \citep{HeI_in_Be}.\\

\begin{figure*}
\centering
\includegraphics[scale=0.7]{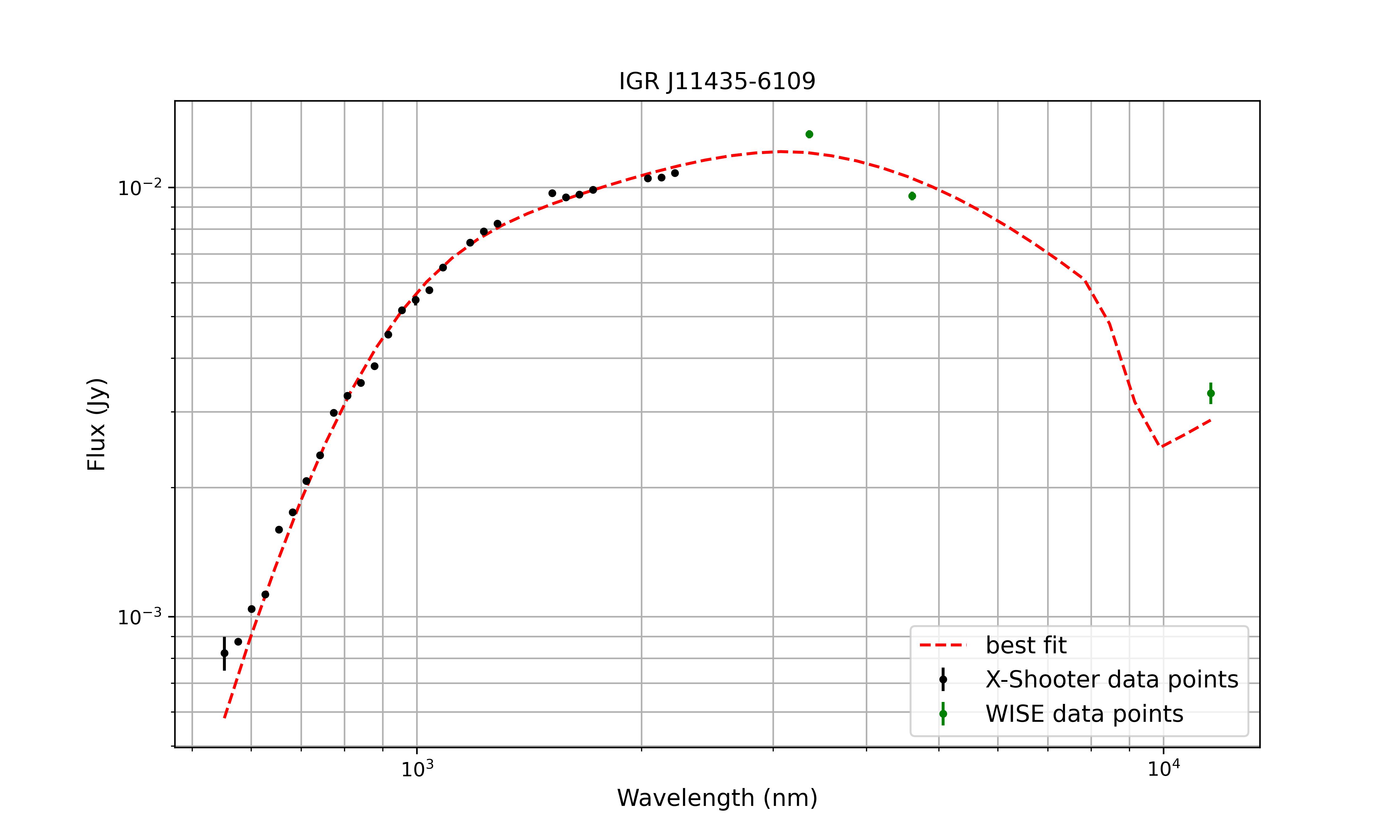}
\caption{Fitting of the SED of IGR J11435-6109, the data points are from the X-Shooter spectrum and \emph{WISE}}
\label{fig:sed_11435}
\end{figure*}

The fitting of the SED (see Fig.\ref{fig:sed_11435}) is done with an isotropic black body representing the companion star and kept at a temperature of 28,000 K, to which we added a disk around the star, with a flux of the form:
\[F_{\nu}=\int_{R_{min}}^{R_{max}} B_{\nu}(T_d(R)) \frac{2\pi R \ \cos{i}}{D^2}\ dR \]
where $B_{\nu}$ is the Planck function, $R$ the distance from the center of the disk, $D$ the distance of the source, and $T_d(R)$ the temperature inside the thin disk, which depends on the radius as \[T_d(r)= K \ T_{max} \ r^{-\frac{3}{4}}\ {(1-r^{-\frac{1}{2}})}^{\frac{1}{4}}\]
where $K \approx 2.04$, $r=\frac{R}{R_{max}}$ and $T_{max}$ the maximum temperature inside the disk \citep{accretion_process}.\\

This disk adds 3 parameters to the model related to the maximum temperature,  the internal and the external radius. 
The two dimensions of the disk are highly correlated so they could not be constrained. We found a maximum disk temperature of $2070 \pm 650$ K, an extinction $A_V = 5.8 \pm 0.3\ mag$ and taking into account the distance uncertainty we find a stellar radius between 7.3 and 11.7 $R_{\odot}$, typical of a main-sequence B0 star \citep{radius_stars_book}.\\

For this source only, we also have access to the periods of the system. A period of 161.76 s which corresponds to the rotation of a neutron star (NS) was computed by \citet{spin_period_11435} using a Fourier transform. A longer period of $52.46 \pm 0.06$ days corresponding to the orbital period of the NS was found by \citet{orbit_period_11435} who fitted a sine wave to the light curve. Note that the variation of luminosity during the orbit is due to its eccentricity. Using a mass-luminosity relation of the type $L \propto M^{7/2}$ \citep{masse_lum}, we can estimate a mass of $M=15.5 \pm 1.2\  M_{\odot}$ (typical of an early B star). Along with the period and Kepler's 3rd law, this mass gives an orbital separation of \(a=151 \pm 3 R_{\odot} = 0.70 \pm 0.01\ AU \), where the NS mass is estimated around $1.4 \pm 0.1 M_{\odot}$ \citep{Kiziltan_2013}. From \citet{orbit_period_11435} and from the observation date of the X-Shooter spectrum, we also found the orbital phase \(\phi_{obs}=0.072 \pm 0.001\), meaning the spectrum was taken at peak X-ray luminosity, when the NS was interacting with the decretion disk of the Be.\\

Furthermore, we can estimate at which distance from the center of the star each line is emitted. From our fits of the double peaked lines, we have $v_d \sin{(i)} = \delta v_p/2$ (where $v_d$ is the speed of the emission zone in the disk) and Kepler's 3rd law can be rewritten as $R_d = \frac{GM}{v_d^2}$, assuming the emission comes from a thin disk in Keplerian motion. Unfortunately the inclination of the decretion disk is not known, so we can only estimate the maximal radius of emission: 
\[ R_{d,max}=  \frac{GM}{ \left(v_d \sin{i}\right)^2} \]

We took the average value of $\delta v_p$ for each hydrogen series and computed the corresponding value of $R_{d,max}$. All the values were compatible with one another so we took the overall average from all the hydrogen lines and found $R_{d,max}= 116 \pm 12 R_{\odot} = 0.53 \pm 0.05\ AU$. This radius is smaller than the orbital separation by 25\%, which means that the NS can only interact with this portion of the disk when it comes close enough along its eccentric orbit.\\

In conclusion, this source is confirmed to be a BeHMXB with a companion star of type B0.5Ve. The compact object (NS) has an eccentric orbit that extends outside the hydrogen-emitting zone of the decretion disk.

\begin{table}
\caption{Double peaked emission lines of IGR J11435-6109}
\begin{threeparttable}
\centering
\begin{tabular}{cccccc}
\hline \hline 
Line    & $\lambda_0$ & $\delta v_p$     & $V$\slash$R$         & $v_{rad}$  \\
        & nm       & $km/s$ &     &     $km/s$ \\
\hline
Balmer  & 656.27   &  339.9  &  1.2  &   25  \\
Paschen & 850.24   &  296  &  1.2  &  -75     \\
        & 854.53   &  285.8  &  1.2  &  -37     \\
        & 859.83   & 326.2  &  1.1  &   12        \\
        & 875.04   & 286.6  &  0.8  &   -4   \\
        & 886.28   & 328  &  1.5  &   15     \\
        & 901.53   &  341  &  1.4  &   10   \\
        & 1281.80       &  321  &  1.4  &   22            \\
Brackett & 1519.18 &  317  &  1.3  &   14 \\
    & 1526.05 &  353  &  1.9  &   32 \\
    &1534.17 &  313.6  &  1.2  &    5 \\
    & 1543.89 &  315.3  &  1.3  &   13 \\
    & 1555.64 &  317 &  1.4  &    8 \\
    & 1588.054 &  311  &  1.28 &    6 \\
    &   1610.93       &  315  &  1.4  &    2   \\
        &    1640.68     &  303  &  1.5  &   10       \\
        &1680.65&    318  &  1.5  &    7          \\
\hline
FeI  & 892.66 &  383  &  1.7  &   17 \\
FeII    & 1050.15 &  336.0  &  1.4  &   15 \\
FeII    &   1086.26 &  309  &  1.1  &   0 \\
CI  &1689.5 &  347  &  1.0  &   55 \\
MgII &2137.4 &   346  &  0.8  &  -62 \\
MgII & 2143.7 &   296  &  0.8  &  -18 \\ \hline 
\end{tabular}
\begin{tablenotes}
\item $V$\slash$R$ is the ratio of intensities of the blushifted peak over the redshifted peak
\end{tablenotes}
\end{threeparttable}
\label{table:11435_H}
\end{table}

\begin{table*}
\begin{threeparttable}
\caption{Shell profile lines of IGR J11435-6109}
\label{table:11435_auto}
\centering
\begin{tabular}{cccccccc}
\hline \hline 
 &  & \multicolumn{3}{c}{Emission} & \multicolumn{3}{c}{Absorption} \\
\hline
 Line    &   $\lambda_0$     & FWHM    & Amplitude    & $v_{rad}$  & FWHM     & Amplitude     & $v_{rad}$   \\
     &  nm      &   $km/s$        &  A.U. &  $km/s$ &   $km/s$        &  A.U. &  $km/s$ \\
\hline
  $H_{\beta}$   &    486.13    &  257       &   0.2      &    7      &   152       &    -0.1      &    11      \\
   HeI  &  706.51      &  200       & 0.2       &  68        &  165        &    -0.1      &    58      \\
  HeI   &1083.03 &   258      &   5.2      &  11        & 165         &    -2.7       &     5     \\
  HeI   &  1700.24      &   235      &   1.5      &    15      &   189       &   -0.9       &   14    \\  \hline 
\end{tabular}
\begin{tablenotes}
\item The amplitudes are given in arbitrary units to compare the emission and absorption features
\end{tablenotes}
\end{threeparttable}
\end{table*}

\begin{table}
\caption{Simple lines of IGR J11435-6109}
\begin{threeparttable}

\centering
\begin{tabular}{cccccc}
\hline \hline 
Line    & $\lambda_0$ & FWHM     & EQW     & Flux         & $v_{rad}$  \\
        & nm       & $\AA$ &    $\AA$ &  $erg/s/cm^2$     &   $km/s$ \\
\hline
HeI & 587.56 &  2.1  &  0.65  & -5.5e-16  & 14 \\
HeI & 667.81 &  2.7  &  0.44  & -4.7e-16  & 27 \\
 FeI   &  890.59 &  2.6  & -0.05  &  9e-17  &  5     \\
  CI  &     1068 & 11  & -0.99  &  1.7e-15  & 48    \\
 NI   &   1246.12    &   15    & -0.09      &   2.3e-15    & -111      \\
  NI  & 1246.96      &   23    & -0.20   &   2.1e-15    &  25    \\
  FeI & 1529.9 & 6.2 &  -0.18 &1.9e-16  & -60 \\
   MgI &    1574.07 & \O    &  $<0$     &  $>0$     &  \O     \\
   FeII &  1678.1     &   \O    & $<0$      &   $>0$    & \O       \\
 FeI   &     1696.9 &  7.8  & -0.36  &  3.7e-16  & -75   \\
 FeI & 2204.28 &  3.22  & -0.22  &  1.47e-16  &  6 \\\hline 

\end{tabular}
\begin{tablenotes}
\item A negative value of EQW indicates an emission feature, and a positive value an absorption feature
\end{tablenotes}
\end{threeparttable}
\label{table:11435_simple}
\end{table}

\subsubsection{IGR J12489-6243}
\paragraph{\emph{Previous results}}
This source was discovered by \citet{Bird_2009}. Afterwards, 2 sources detected by \emph{Chandra} were candidates as a possible counterpart, since the location given by \emph{INTEGRAL} is not precise enough \citep{Tomsick_2012}. The authors finally chose the source with the hardest X-ray emission as the true counterpart and concluded it was either a CV or a HMXB. A study by \citet{Fortin_2018} was also done in infrared and the source was found to be a CV with K or M companion star, despite the lack of molecular absorption in the spectrum. They explained this peculiarity by saying that the white dwarf accreted a portion of the atmosphere of the low mass star. The molecules that can usually be seen in the atmosphere of those stars have hence been depleted in large enough quantities so that they do not appear on the spectrum.\\

\paragraph{\emph{Our results}}
Our X-Shooter spectrum is characterized by the presence of hydrogen lines in emission as well as 8 HeI lines, which indicates a relatively helium-rich environment (see Table \ref{table:12489_H}). We also found numerous neutral metal lines in absorption, which hints towards a low mass star  (see Table \ref{table:12489_metals}), but none of the CO absorption bands in the region 2.2 - 2.5 $\mu m$ are detected in the spectrum, which could have been used to constrain the luminosity class by comparing their strength to other metallic lines. Instead we focused on 3 absorption lines of MgI and AlI in the H-band and used their equivalent widths to determine the spectral type. According to \citet{Meyer_1998}, those lines are compatible with a subgiant between K0IV and K2IV.\\

\begin{figure*}
\centering
\includegraphics[scale=0.55]{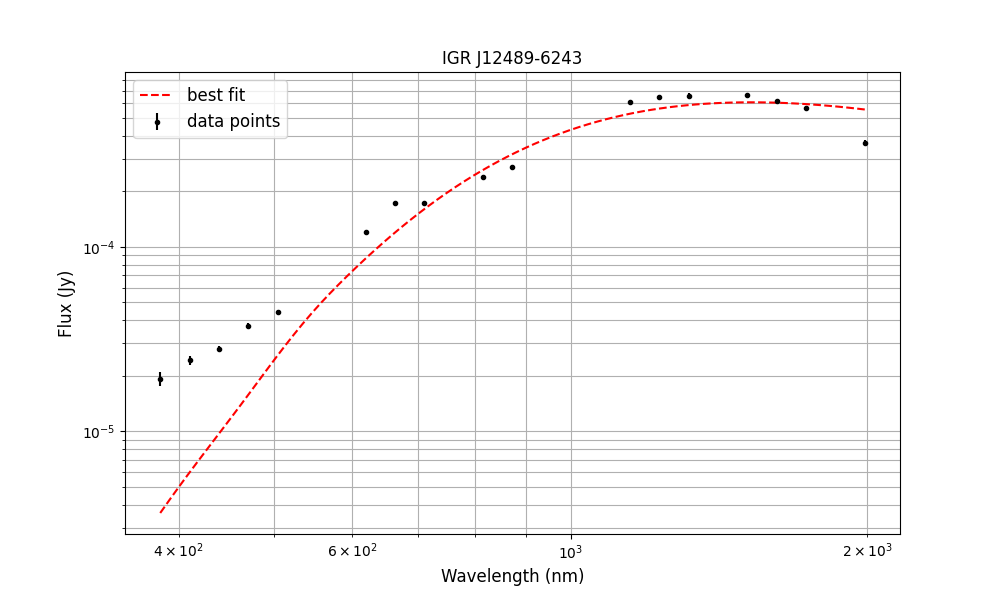}
\caption{Fitting of the SED of IGR J12489-6243, the data points are from the X-Shooter spectrum}
\label{fig:sed_12489}
\end{figure*}

We then tried to confirm this finding by fitting the continuum with a single black body. Unlike the 2 previous sources, the temperature of the star is not well constrained so we made it a free parameter of the model. Unfortunately during the fitting process the interstellar absorption parameter is correlated to the temperature, so we had to keep $A_V$ and $E(B-V)$ as fixed parameters.
The reddening given by the dust reddening map of \citet{Schlafly_2011} is $E(B-V)=12.8$, which seemed too high for a relatively nearby source (distance of 2.5 kpc). When applied to the spectrum this reddening was also found to be inaccurate, since no black body could fit the un-reddened spectrum. Therefore we instead decided to use a method based on the Balmer decrement. Indeed the intensity ratio between the $H_{\alpha}$ and $H_{\beta}$ emission lines can be linked directly to the reddening (see appendix), and for a ratio of 8.64 we found a reddening of $E(B-V) = 2.79$, which was used for the final SED fitting (see Fig.\ref{fig:sed_12489}).\\

After taking this fixed value of the reddening, the star temperature fitting the continuum is found to be 5046 $\pm$ 206 K, which is coherent with an early K star in the atlas of \citet{Meyer_1998}. Taking into account the uncertainty on the distance of the source, the radius is $4.4^{+1.8}_{-1.0}\ R_{\odot}$ and the corresponding luminosity is $11^{+15}_{-6}\ L_{\odot}$, which is compatible with an evolved star of type subgiant or giant \citep{radius_stars_book}. This confirms that the companion star is of type K0IV-K2IV, but further investigations could be done to constrain the luminosity if the distance is known more precisely.\\

Moreover, on the fit in Fig.\ref{fig:sed_12489} we clearly see a growing discrepancy between the data and the blackbody emission when going towards higher energy, especially between 380 and 500 nm. This excess in flux in the UV could be explained by another source of light at higher energy that adds up to the flux of the blackbody. Such a source could be the accretion disk or the white dwarf of the CV system that emits some of its light in UV . This contamination from another black body, as well as the weakness of some molecular lines was also observed by \citet{harrison_2004_CV} and \citet{CV_UV_disk}. Unfortunately the fitting process doesn't converge when we add more parameters so the origin of this discrepancy remains unclear.\\
 
We also note the presence of 3 CaII lines in emission, that we had to disentangle from the nearby Paschen lines. This single-ionised metal is usually present in absorption in K-M star but we find it here in emission, which means that this element is in the accretion disk of the compact object, indicating that it was stripped from the stellar atmosphere. This Calcium triplet is sometimes observed in CV and is a marker of the cool outer regions of the accretion disk, similarly to what is observed in AGNs \citep{CaII_CV}.
This corroborates the explanation of \citet{Fortin_2018} concerning the lack of molecular absorption in the spectra: both the molecules usually present in the star and the CaII were stripped from the atmosphere.\\ 

We can then confirm that this source is a Helium-rich CV and we constrain the companion star to a K0-K2 subgiant with an atmosphere which has been stripped of molecules and other elements like Ca. The source cannot be a polar since we observe emission lines coming from a disk, the source is then a Intermediate Polar (IP). This is confirmed by the X-ray luminosity of $4 \times 10^{32}$ erg/s, which is typical of a high-luminosity IP \citep{luminosity_IP}.

\begin{table}
\caption{Emission lines of IGR J12489-6243}
\label{table:12489_H}
\centering
\begin{tabular}{cccccc}
\hline \hline 
Line    & $\lambda_0$ & FWHM     & EQW     & Flux         & $v_{rad}$  \\
        & nm       & $\AA$ &    $\AA$ &    $erg/s/cm^2$   &   $km/s$ \\
\hline
Balmer  & 486.13   & 7.4      & -13     & 6.7e-16      & 77     \\
        & 656.27   & 10.5     & -39     & 4.86e-15     & 24     \\
Paschen & 850.24   & 7.0      & -3.2    & 3.5e-16      & -183   \\
        & 854.53   & 17       & -14     & 1.4e-15      & -56    \\
        & 859.83   & 15       & -7.8    & 7.6e-16      & 18     \\
        & 866.50   & 17.3     & -20.0   & 1.74e-15     & -17    \\
        & 875.04   & 17.5     & -14.6   & 1.32e-15     & -2     \\
        & 886.28   & 16.1     & -16.8   & 1.50e-15     & 6      \\
        & 901.53   & 14.6     & -15.1   & 1.27e-15     & -8     \\
        & 922.97   & 18.5     & -21.6   & 2.07e-15     & -50    \\
        & 954.62   & 6.3      & -8.0    & 7.0e-16      & 80     \\
        &   1093.81       &12.0&    -7.8       &      1.26e-15      &      44  \\
        & 1281.80         &   13.0       &    -10.1     &      1.41e-15        &  56      \\
Brackett&   1610.93       &   27.5       &  -7.1       &5.0e-16&  48      \\
        &    1640.68      &    25.3      &    -9.8     &   6.46e-16           &  62      \\
        &1680.65&     25.0     &   -10.1      &      6.67e-16        &  34      \\
        &  1736.0        &   25.8       &   -15.3      &    8.59e-16          &    77 \\
        & 2166.1  & 17.2  & -21  &        6.40e-16 & -40\\
HeI &   667.81     &   5.8     &  -3.9     &  4.158e-16     & 70     \\
        &  706.51      &    6.06    &   -3.0    &  3.36e-16     &  85    \\
        &   728.13     &  3.1       &  -1.8     &   1.0e-15    &  100    \\
        &    1083.03    &   16.5     &  -23.9     &  3.78e-15     & 11     \\
        &  1279.05      &  20      &  -5     &  6e-16     &  8    \\
        &   1700.24     &  13      &  -2     & 1e-16      &  83    \\
        &   2058.69     &  24      &   -24    &  6.0e-16     &  -39    \\
CaII & 849.80 &  5.1  &  -2.0  &  2.7e-16  & -56 \\
    & 854.20 & 2.8  &  -1.2  &  1.5e-16  & -60 \\
   & 866.21 &  1.5  &  -0.3  &  4.3e-17  & -59 \\\hline 

\end{tabular}
\end{table}

\begin{table}
\caption{Absorption lines of IGR J12489-6243}
\label{table:12489_metals}
\centering
\begin{tabular}{cccccc}
\hline \hline 
Line    & $\lambda_0$ & FWHM     & EQW     & Flux         & $v_{rad}$  \\
        & nm       & $\AA$ & $\AA$ & $erg/s/cm^2$ & $km/s$ \\
\hline

    FeII &  942.84    &  2.1     &   2.0    &   -3.5e-16    &  -41       \\
    SiI    &  1060.34     &  2.2     &   2.6    &  -3.6e-16    &     12  \\
    MgI    &  1182.81     &  2.5     &   0.6    &   \O    &  -129     \\
    MnI    &  1290.0     &  1.8     &  0.4     &  -5e-17     &  17\\    
    MgI    &  1502.49     &  5     &   0.8    &   -7e-17    &   -107    \\
    MgI    &  1576.58     &  7.5     &  1.9     &  -1.5e-16     &  -64     \\
    AlI    &   1671.89    &    6   & 0.8      &    -5e-17   &   -54    \\
    AlI    &   1675.0    &  5.5     & 1.4    &  -9.3e-17     &   -65    \\
    MgI    &   1710.86    &5&    1.5   &  -8e-17     &  -92     \\
    NaI    &  2205.32      &  6     &   3    &  -6e-17     &    -41   \\ \hline 
\end{tabular}
\end{table}

\subsection{Peculiar CI emission in the Be stars}
\begin{figure*}
    \centering
    \subfloat[IGR J10101-5654]{
    {\includegraphics[scale=0.32]{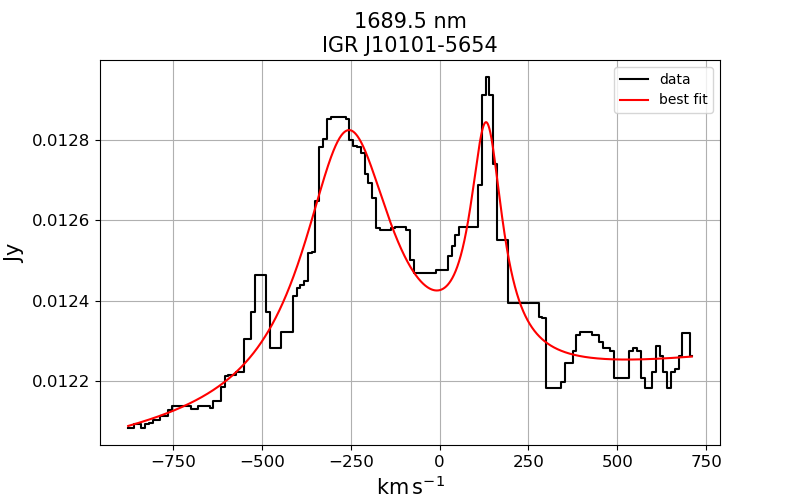} }}%
    \subfloat[IGR J11435-6109]{
    {\includegraphics[scale=0.32]{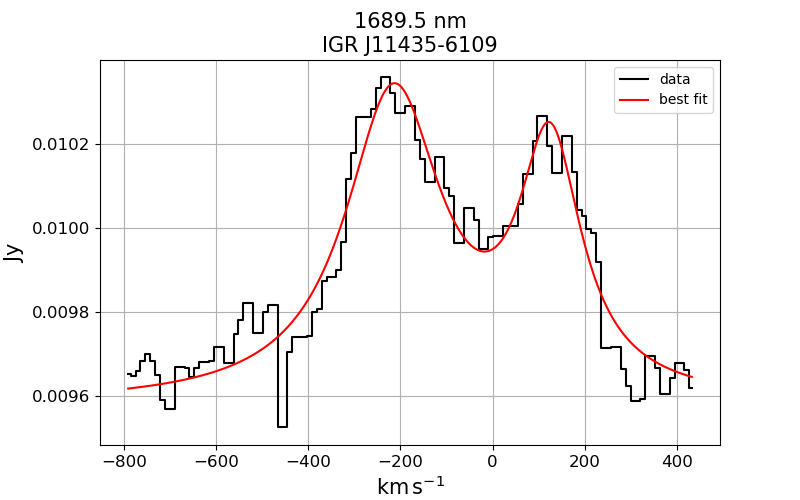} }}%
    \hspace{0mm}
    \caption{CI double-peaked emission with fit, the spectrum of IGR J10101-5654 is smoothed around this feature because of local atmospheric features}%
    
\label{fig:CI_emission}

\end{figure*}

In both Be we studied, we observed a double-peaked emission of neutral carbon at 1689 nm, although it is uncertain for IGR J10101-5654. The lines can be seen in Fig. \ref{fig:CI_emission} along with the fits, and look similar in terms of width, amplitude and shape. The CI line of IGR J10101-5654 is strongly affected by sky absorption but becomes clearly visible after a median filter is applied to smooth the spectrum near the line.\\

\citet{Chojnowski_2015} observed this line in 18 to 26\% of the 238 Be studied, and our two sources have quite similar spectrum overall, which is why it would not be too surprising to find this peculiar emission in both. Numerous neutral metal were observed in emission, notably other CI emission lines around 1068 nm, and FeI emissions. The presence of FeI was also found by \citep{Chojnowski_2015} to be associated with the CI 1689 nm emission, in particular FeI 1529 nm, which is indeed observed for IGR J11435-6109.\\

\section{Conclusions}
Thanks to the X-Shooter spectra, we were able to constrain the nature of 3 sources and find many relevant physical parameters concerning their close environment:

   \begin{enumerate}
      \item \emph{IGR J10101-5654} was previously thought to be a supergiant B[e] but our investigation showed no real evidence for this source to be one, in particular because of the lack of reliable detection of forbidden lines ($S/N \approx 2.5$). Using a star atlas we found the companion star to be compatible with a B0.5Ve with peculiar carbon emission, thus a BeHMXB system. The double-peaked lines also confirm the presence of a decretion disk. The flux variation in both X-ray and optical over time could hints towards a variability of the source.
      \item \emph{IGR J11435-6109} has many similarities to IGR J10101-5654, with which it shares the same companion star atlas compatible with a B0.5Ve with the presence of double-peaked lines. We could also estimate the orbital separation between the NS and the companion star, as well as the maximal extension of the zone of hydrogen emission in the decretion disk.
      \item \emph{IGR J12489-6243} was confirmed to be a CV. The companion star was also constrained to be a subgiant K0IV-K2IV by using a stellar atlas in the H-band and a fit of the SED of the spectra. We did not find any molecular bands in the spectra so this companion star seems to have been stripped from some of its photosphere. This is confirmed by the presence of CaII in emission, i.e. in the outer accretion disk, which also indicates that the source is an IP. 
   \end{enumerate}
    
A further study of the sources could be done in the future, for instance by comparing spectra of the same source taken at different orbital time. In the case of Be stars this could help us to understand how the shape of the decretion disk changes with time by following the evolution of the double peaks in the spectrum, in particular the ratio in intensity between the two peaks ($V$\slash$R$) and the inter-peak distance ($\delta v_{p}$) which were already measured here.\\

\section*{Acknowledgements}
Based on observations made with ESO Telescopes at the La Silla Paranal Observatory under program ID 0102.D-0918(A).\\
SC is grateful to the Centre National d’Etudes Spatiales (CNES) for the funding of MINE (Multi-wavelength INTEGRAL Network).\\
This work has made use of data from the European Space Agency (ESA) mission {\it Gaia} (\url{https://www.cosmos.esa.int/gaia}), processed by the {\it Gaia} Data Processing and Analysis Consortium (DPAC, \url{https://www.cosmos.esa.int/web/gaia/dpac/consortium}). Funding for the DPAC has been provided by national institutions, in particular the institutions participating in the {\it Gaia} Multilateral Agreement.\\
JAT acknowledges partial support from the National Aeronautics and Space Administration (NASA) through Chandra Award Number GO9-20045X issued by the Chandra X-ray Observatory Center, which is operated by the Smithsonian Astronomical Observatory under NASA contract NAS8-03060.
\section*{DATA AVAILABILITY STATEMENT}
The data underlying this article are available in the article.

\bibliographystyle{mnras}
\bibliography{reference}

\newpage
\begin{appendix}

\FloatBarrier
\section{Balmer decrement}
The observed intensity of a line at a wavelength $\lambda$ depends on the optical depth $\tau_{\lambda}$ and its intensity without interstellar absorption $I_{0}$ as: \[I = I_{0}\ e^{-\tau_{\lambda}} \]
In the UV, \citet{Calzetti_1994} found that the optical depth is linked to the reddening $E(B-V)$ with the relation:
\[ k_{\lambda}=\frac{1.086\ \tau_{\lambda}}{E(B-V)}\]
where $k_{\lambda}=\frac{A_{\lambda}}{E(B-V)}$ and $A_{\lambda}$ is the extinction.\\
We can replace the optical depth in the first relation to get:
\[I = I_{0}\ \exp{\left(\frac{-k_{\lambda}\ E(B-V)}{1.086}\right)}\] which can be re-written as \(I = I_{0}\ 10^{- 0.4\ k_{\lambda}\ E(B-V)} \).
The absorbed intensity of $H_{\alpha}$ and $H_{\beta}$ are known, we can then invert the last relation to find $E(B-V)$:
\[ E(B-V)= \frac{\log{\left( \frac{I_{0\ \alpha}}{I_{0\ \beta}}\right)}-\log{\left(\frac{I_{\alpha}}{I_{\beta}}\right)}}{0.4\ (k_{\alpha}-k_{\beta})}\]
The relative intensity of the lines without absorption are known physical parameters: \(\frac{I_{0\ \alpha}}{I_{0\ \beta}}=2.86\) \citep{balmer_ratio}, moreover \citet{seaton} found \(k_{\alpha}-k_{\beta}=-1.25\) for our galaxy.\\
This gives us a direct relation between the intensity of $H_{\alpha}$ and $H_{\beta}$, and the reddening:
\[ E(B-V)=2\ \log{\left( \frac{I_{\alpha}}{2.86\ I_{\beta}}\right)}\]

\end{appendix}

\bsp	
\label{lastpage}
\end{document}